\documentclass[conference]{IEEEtran}
\IEEEoverridecommandlockouts

\usepackage[T1]{fontenc}
\usepackage[utf8]{inputenc}
\usepackage{amsmath,amssymb}
\usepackage{array,booktabs,makecell,tabularx}
\usepackage{graphicx}
\usepackage{tikz}
\usetikzlibrary{positioning}
\usepackage{pgfplots}
\usepackage{url}
\usepackage{enumitem}
\usepackage{microtype}
\pgfplotsset{compat=1.18}

\newcolumntype{Y}{>{\raggedright\arraybackslash}X}
\newcommand{\accfield}[2]{\textbf{#1} & #2 \\}

\begin{document}

\title{Designing Intelligent Enterprise Agents:\\
A Capability-Aligned Multi-Agent Architecture}

\author{\begin{tabular}{c}
John deVadoss\\
InterWork Alliance\\
Washington DC\\
johnd@ieee.org
\end{tabular}}

\maketitle

\begin{abstract}
Enterprise interest in multi-agent systems has shifted from generic software agents to large-language-model (LLM) based intelligent agents that plan, use tools, maintain contextual memory, inspect intermediate results, collaborate with other agents, and sometimes act in systems of record. This paper revises the enterprise architecture thesis around a design-first claim: governance is necessary, but it cannot be the primary organizing abstraction. The primary abstraction must be agent design: capability boundaries, autonomy allocation, interaction protocols, tool and data authority, state and memory design, verification design, and human interaction design. We propose CEAD -- Capability-Aligned Enterprise Agent Design -- a reference architecture for LLM-based intelligent enterprise agents. CEAD uses service-oriented architecture (SOA) as an exemplar for contracts, registries, loose coupling, and policy-aware integration, while explicitly rejecting the idea that services are agents. It treats microservices as a cautionary precedent: decomposition without design discipline produces distributed complexity, cost, operational fragility, and agent proliferation. We define enterprise intelligent agents, introduce the Agent Capability Contract (ACC) as a design artifact rather than a governance artifact, and provide design principles for multi-agent decomposition. We evaluate CEAD over 10,000 enterprise tasks and compare a prompt-first mono-agent, a role-based micro-agent swarm, SOA-brokered agents, a governance-first but design-poor agent grid, and the proposed CEAD architecture. CEAD achieves 70.6\% safe success versus 45.2\% for a mono-agent baseline, 23.1\% for an ungoverned micro-agent swarm, 58.8\% for SOA-brokered agents, and 50.8\% for a control-heavy design-poor grid. The results support the conclusion that design quality is the first-order enterprise concern; governance, security, policy, audit, and assurance should support and enforce good design rather than substitute for it.
\end{abstract}


\section{Introduction}
The term agent has a long history in artificial intelligence and distributed systems. Wooldridge and Jennings characterized intelligent agents in terms such as autonomy, social ability, reactivity, and proactiveness \cite{wooldridge1995}. The current enterprise wave, however, is narrower and more operational: organizations are building LLM-based agents that interpret natural-language intent, reason over context, call tools and APIs, browse or retrieve information, produce intermediate plans, delegate subtasks, and sometimes take actions in systems of record. Recent surveys describe LLM-based autonomous agents as systems that build on language-model reasoning, planning, memory, tool use, and evaluation mechanisms \cite{wang2024}; multi-agent surveys emphasize profiling, communication, coordination, skill acquisition, and benchmarks for LLM-based multi-agent systems \cite{guo2024}.

This paper therefore explicitly scopes intelligent enterprise agents as LLM-mediated software actors with bounded autonomy, enterprise identity, tool access, governed memory, and accountable actions. The scope excludes simple API services, deterministic workflow steps, stateless chatbots, and loosely packaged prompts unless they possess the agentic properties defined in Section~\ref{sec:definition}. It also excludes unrestricted autonomous systems: in enterprises, intelligent agents must remain bounded by policy, data classifications, identity, audit, release controls, and human approval for high-risk actions.

The architectural challenge is familiar. Enterprise architecture has already lived through component proliferation. SOA provided useful ideas -- contracts, service registries, policies, interoperability, reuse, and explicit interaction semantics -- but also suffered when implemented as technology fashion rather than design discipline. Microservices later emphasized independently deployable services organized around business capabilities, but widespread misunderstanding produced distributed monoliths, excessive operational overhead, and service sprawl \cite{lewis2014,fowler2015,su2024}. Intelligent agents could repeat this pattern at higher risk: instead of too many services, enterprises may create too many prompt-wrapped micro-agents with overlapping authority, opaque memory, inconsistent tool scopes, and untestable emergent behavior.

The contributions of this paper are fourfold:
\begin{enumerate}
  \item It defines the scope of LLM-based intelligent enterprise agents and distinguishes agents from services, microservices, bots, and ordinary workflows.
  \item It evaluates SOA as an architectural exemplar for contracts, registries, loose coupling, and interoperability while identifying where SOA is insufficient because agents are autonomous, probabilistic, stateful, and goal-directed.
  \item It proposes CEAD, a design-first, capability-aligned architecture in which governance supports, enforces, and audits design decisions rather than replacing design.
  \item It implements an evaluation of architectural principles, including agent proliferation, design-versus-governance mismatch, ablations, and high-risk/adversarial stress tests.
\end{enumerate}

\subsection{Clarifying Questions and Working Assumptions}
A concrete enterprise deployment should answer design questions before implementation: What business capability is the agent responsible for? Who owns that capability? What decisions may the agent make, and which actions must remain human-owned? What state and memory does the task require? Which services, tools, data products, and external agents are legitimate capabilities for the agent to use? What interaction topology is justified -- a single supervised agent, supervisor-plus-specialists, peer collaboration, or deterministic workflow plus agentic assistance? Governance questions then support those design decisions: what data classes may be accessed, what approvals are required, which identity and policy systems are authoritative, what evaluation thresholds are release gates, and how audit and incident response are implemented?

Because those answers are enterprise-specific, this paper assumes a cross-industry enterprise with existing services and microservices, moderate regulatory exposure, centralized identity, heterogeneous data products, a mixture of internal and external LLM providers, and a requirement that high-risk actions be observable, reversible where possible, and subject to human approval.

\section{Background and Related Work}
\subsection{From Intelligent Agents to LLM Agents}
Classical agent theory is broader than LLM-based agents. Agent-oriented software engineering treated agents as autonomous systems that interact to satisfy design objectives \cite{jennings2000}. LLM-based agents differ in implementation: the model often acts as a planner, interpreter, router, or critic, while tools provide perception and action. ReAct demonstrated interleaving reasoning traces with actions so that language models can interact with external sources and update plans \cite{yao2023}. Toolformer showed that language models can learn when and how to call external APIs \cite{schick2023}. Generative Agents combined memory, reflection, and planning to simulate believable social behavior \cite{park2023}. AutoGen demonstrated multi-agent LLM applications built through conversational interactions among configurable agents, humans, and tools \cite{wu2024}.

Benchmarks show both promise and fragility. AgentBench evaluates LLMs as agents across multiple interactive environments \cite{liu2023}. WebArena reports that a best GPT-4-based web agent achieved 14.41\% end-to-end success on realistic web tasks versus 78.24\% for humans \cite{zhou2024}. GAIA emphasizes real-world questions requiring reasoning, multimodality, browsing, and tool use, with human performance far exceeding early GPT-4-with-tools performance \cite{mialon2024}. SWE-bench evaluates whether models can resolve real GitHub issues by producing patches \cite{jimenez2024}. These results motivate enterprise caution: high-level language ability does not imply reliable multi-step action.

\subsection{SOA, Microservices, and Enterprise Architecture}
OASIS defines the SOA Reference Model as an abstract framework for understanding entities and relationships in a service-oriented environment, not as a concrete technology stack \cite{oasis2006}. This matters for agents. SOA is valuable as a design exemplar because it emphasizes visibility, interaction, real-world effect, contracts, policies, and shared semantics. Yet a service is generally a passive capability invoked through a contract; an intelligent agent is an active goal-directed actor that chooses actions, selects tools, maintains context, and may communicate with other agents.

Microservices refined service decomposition around business capabilities, independent deployment, decentralized governance, and automation \cite{lewis2014}. The caveat is that microservices also introduced a premium in complexity, cost, and operational risk when adopted prematurely or overzealously \cite{fowler2015}. Recent multivocal literature on returning from microservices to monoliths identifies cost, complexity, scalability, performance, and organization as recurring reasons for backtracking \cite{su2024}. Agent architectures should learn from this history: do not decompose into many small agents unless the resulting autonomy, ownership, evaluation, and coordination boundaries are justified.

\subsection{Protocols, Governance, and Security}
Interoperability standards are emerging for agentic systems. The Model Context Protocol (MCP) standardizes how applications provide context, tools, and data to LLMs, and its specification highlights security and trust considerations such as user consent, control, and data privacy \cite{mcpintro,mcp2025}. Agent2Agent (A2A) aims to enable secure communication and interoperability among agents built with different frameworks or by different vendors \cite{a2a2025}. These protocols are useful carriers, but they do not by themselves solve authorization, policy, data classification, provenance, or accountability.

Governance references provide enterprise context. NIST's Generative AI Profile extends the AI Risk Management Framework to generative AI risks \cite{nist2024}. ISO/IEC 42001 specifies requirements for establishing, maintaining, and improving an AI management system \cite{iso42001}. OWASP's Top 10 for Agentic Applications identifies critical risks for autonomous and agentic AI systems \cite{owasp2025}. Indirect prompt injection work highlights that LLM-integrated applications can be manipulated through external content because models may fail to distinguish instructions from data \cite{yi2023}. The architecture proposed here treats these concerns as first-class design constraints.

\section{Definition: Intelligent Enterprise Agents}
\label{sec:definition}
\subsection{Definition}
An intelligent enterprise agent is a software actor that satisfies the following conditions:
\begin{enumerate}
  \item \textbf{Goal-directed intent handling:} it accepts an objective, not merely a procedure call.
  \item \textbf{LLM-mediated cognition:} it uses an LLM or comparable foundation model for interpretation, planning, reasoning, critique, communication, or tool routing.
  \item \textbf{Tool-mediated action:} it can call enterprise services, APIs, databases, search, calculators, code interpreters, workflow engines, or other agents.
  \item \textbf{Bounded autonomy:} it can choose among actions within explicit authority, policy, and approval boundaries.
  \item \textbf{State and memory:} it may maintain task state, short-term context, and optionally long-term memory governed as enterprise data.
  \item \textbf{Observable accountability:} it has an owner, identity, trace, evaluation record, cost record, and escalation path.
  \item \textbf{Uncertainty handling:} it can detect ambiguity, request clarification, seek verification, or escalate.
\end{enumerate}

The definition intentionally requires both intelligence and enterprise accountability. A prompt that summarizes a document is not necessarily an agent. A microservice that exposes a REST API is not necessarily an agent. An enterprise agent is created when model-mediated decision-making is combined with tool access and delegated authority.

\subsection{Autonomy Levels}
Enterprises should avoid a binary ``agent allowed/not allowed'' model. We define five autonomy levels:
\begin{itemize}
  \item \textbf{L0 Observe:} retrieve, summarize, classify, or explain; no system changes.
  \item \textbf{L1 Draft:} produce recommendations or drafts; human executes.
  \item \textbf{L2 Prepare:} prepare structured actions; human approves submission.
  \item \textbf{L3 Bounded execute:} execute reversible or low-risk actions under policy and budget.
  \item \textbf{L4 High-autonomy execute:} execute high-impact workflows only with explicit enterprise risk acceptance, continuous monitoring, and strong rollback/escalation mechanisms.
\end{itemize}

\section{SOA as Exemplar, Microservices as Warning}
\subsection{Why SOA Helps}
SOA is a useful exemplar because enterprise agents need many of the same disciplines that services need:
\begin{itemize}
  \item \textbf{Contracts:} agents must publish capabilities, inputs, outputs, constraints, and expected effects.
  \item \textbf{Registries:} other systems must discover what an agent can do and who owns it.
  \item \textbf{Policies:} invocation must be conditioned by identity, data classification, risk, cost, and context.
  \item \textbf{Loose coupling:} agents should interact through explicit contracts rather than hidden prompt dependencies.
  \item \textbf{Observability:} interactions, decisions, and effects must be traceable.
  \item \textbf{Versioning:} agent prompts, models, tools, memories, and policies change and must be release-managed.
\end{itemize}

\subsection{Why SOA Is Insufficient}
The caveat is fundamental: services are not necessarily agents. A service exposes a capability; an agent decides how to pursue a goal. Services are usually evaluated for functional correctness, availability, performance, and contractual conformance. Agents must additionally be evaluated for plan quality, tool selection, grounding, hallucination resistance, prompt-injection resistance, memory safety, delegation safety, and uncertainty calibration. SOA provides governance vocabulary, but not enough behavioral control for probabilistic, context-dependent, model-mediated actors.

\subsection{The Micro-Agent Proliferation Failure Mode}
The most likely enterprise anti-pattern is micro-agent proliferation: creating many small agents because microservices became familiar, because framework demos use role-playing agents, or because prompt wrappers appear cheap. This can create:
\begin{itemize}
  \item duplicated capabilities and unclear ownership;
  \item unnecessary handoffs and long latency chains;
  \item inconsistent tool permissions;
  \item multiple ungoverned memory stores;
  \item poor evaluation coverage because each agent has a small local test but the system has emergent global behavior;
  \item cascading errors, where one agent's confident mistake becomes another agent's input.
\end{itemize}
The design implication is: start with business capabilities and risk boundaries, not with a target number of agents. Decompose only where there is a durable reason, such as separate business ownership, materially different tool authority, distinct data classification, specialized evaluation, or independent release cadence.

\begin{table*}[t]
\caption{Agents, Services, Microservices, Bots, and Workflows}
\label{tab:comparison}
\scriptsize
\centering
\begin{tabularx}{\textwidth}{p{0.13\textwidth}YYYYY}
\toprule
\textbf{Property} & \textbf{SOA service} & \textbf{Microservice} & \textbf{RPA/workflow bot} & \textbf{LLM-based intelligent agent} & \textbf{Enterprise multi-agent system} \\
\midrule
Primary abstraction & Contracted capability & Independently deployable business capability & Scripted task execution & Goal-directed model-mediated actor & Coordinated set of bounded actors \\
Autonomy & Low; invoked by consumers & Low to moderate; service owns capability & Low; follows script & Moderate to high within policy & Distributed autonomy with orchestration or peer protocols \\
Reasoning & External to service & External to service & Rule/script based & LLM-mediated planning, critique, and tool selection & Role-specialized reasoning and delegation \\
State/memory & Usually explicit service state & Service-owned state & Script state & Task context plus governed memory & Shared, isolated, or brokered memory patterns \\
Interface & API/contract & API/event contract & UI or workflow trigger & Intent, tool calls, messages, approvals & Agent messages, task contracts, service/API calls \\
Main risk & Integration mismatch & Distributed operational complexity & Brittle UI automation & Hallucination, tool misuse, prompt injection, overreach & Emergent coordination failure, cascading errors, agent sprawl \\
Primary design artifact & Service contract/SLA & Product ownership and SLO & Runbook/change control & Agent Capability Contract & Capability map, contracts, topology, controls, evals, audit \\
\bottomrule
\end{tabularx}
\end{table*}

\section{Proposed Architecture: CEAD}
CEAD stands for Capability-Aligned Enterprise Agent Design. It is a design-first reference architecture for LLM-based intelligent agents in the enterprise. Its central premise is that multi-agent systems should be decomposed around durable business capabilities, authority boundaries, state boundaries, evaluation boundaries, and ownership boundaries. Governance is still required, but it is a supporting control and assurance plane that enforces the design, detects drift, and provides accountability.

\subsection{Design Thesis}
The enterprise design problem is not ``how many agents should be built?'' but ``what responsibilities, decisions, actions, state, and evidence should be assigned to which bounded actors?'' A governance-first architecture may produce many registered, audited agents that are still badly designed: overlapping capabilities, excessive handoffs, unclear authority, weak evaluation oracles, and duplicated memory. CEAD therefore treats governance as necessary but secondary. The first-order design objects are capability, boundary, autonomy, interaction, state, and evaluation.

\begin{figure*}[t]
\centering
\begin{tikzpicture}[font=\footnotesize, node distance=0.28cm]
  \tikzset{layer/.style={draw, rounded corners, minimum width=0.78\textwidth, minimum height=0.75cm, align=center}}
  \node[layer] (intent) {\textbf{Intent and Outcome Layer}\\ users, applications, events, APIs};
  \node[layer, below=of intent] (design) {\textbf{Agent Design Plane}\\ capability map, ACCs, autonomy allocation, interaction model, evaluation design};
  \node[layer, below=of design] (runtime) {\textbf{Agent Runtime Plane}\\ supervisor, specialist agents, planner, memory/RAG, verifiers, human approval};
  \node[layer, below=of runtime] (capability) {\textbf{Enterprise Capability Plane}\\ SOA services, microservices, data products, MCP servers, A2A peers};
  \node[layer, below=of capability] (control) {\textbf{Supporting Control and Assurance}\\ identity, policy, evals, audit, change control};
\end{tikzpicture}
\caption{CEAD reference architecture. The design plane is primary: capability map, ACCs, autonomy allocation, interaction model, and evaluation design. Governance, security, policy, and audit support the design as a control and assurance sidecar.}
\label{fig:cead}
\end{figure*}
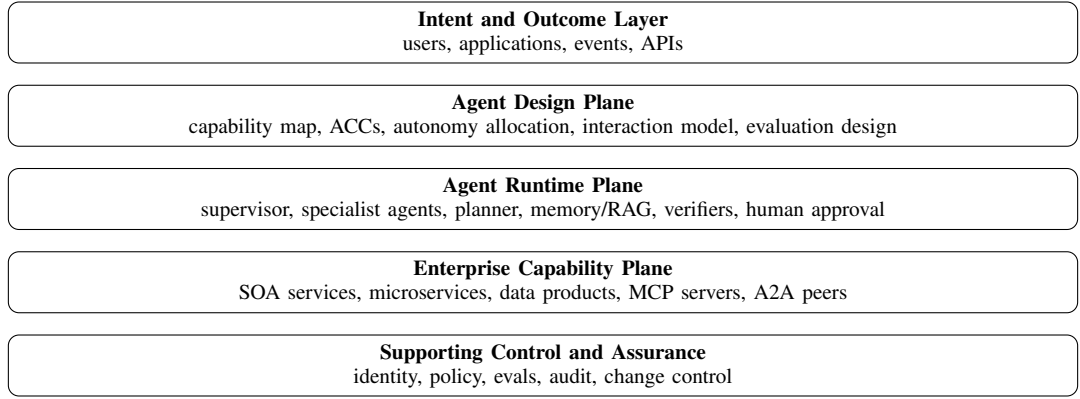

\subsection{Layers}
\textbf{Intent and Outcome Layer.} Users, applications, process triggers, and APIs submit intents. The layer captures user identity, business purpose, tenant, data context, expected outcome, and user-visible evidence. The design question is what outcome the enterprise is willing to delegate, not merely which prompt receives the request.

\textbf{Agent Design Plane.} This is the architectural center of CEAD. It maintains the capability map, agent boundaries, ACCs, autonomy levels, state and memory patterns, interaction topology, verifier strategy, evaluation oracles, and retirement rules. It answers whether an agent should exist, what it owns, what it must not own, which other capabilities it may invoke, and how success is measured.

\textbf{Agent Runtime Plane.} This hosts supervisors, planners, specialist agents, memory/RAG components, deterministic validators, verifiers, tool routers, and human-approval workflows. Runtime components may be implemented with agent frameworks, actor models, workflow engines, or event-driven systems. The runtime must conform to the design plane rather than inventing authority through prompts.

\textbf{Enterprise Capability Plane.} Existing SOA services, microservices, data products, SaaS APIs, knowledge bases, MCP servers, code execution tools, search, and external A2A peers live here. Agents consume these capabilities through explicit contracts and least-privilege adapters; they do not own enterprise capabilities by default.

\textbf{Supporting Control and Assurance Plane.} Identity, authorization, policy, security review, privacy review, model evaluation, red-teaming, release gates, audit, incident response, and change management support the design. This plane provides assurance, not the primary decomposition logic.

\subsection{Agent Capability Contract}
The ACC is the core design artifact. It is analogous to a service contract in SOA, but it is extended for autonomy, model behavior, tools, memory, verification, escalation, and evaluation. The ACC is not introduced as a bureaucratic form; it is the record of the design decision that justifies an agent's existence.

\begin{table*}[t]
\caption{Agent Capability Contract (ACC) as Design Artifact}
\label{tab:acc}
\small
\centering
\begin{tabularx}{\textwidth}{p{0.25\textwidth}Y}
\toprule
\textbf{Contract field} & \textbf{Design purpose} \\
\midrule
\accfield{Business capability and owner}{Names the durable capability, accountable product owner, operating team, funding model, and retirement owner.}
\accfield{Purpose and non-purpose}{Describes what the agent is allowed to do and explicitly what it must not do.}
\accfield{Autonomy level}{Declares L0--L4 action authority, approval requirements, reversibility expectations, and fallback behavior.}
\accfield{Interaction topology}{Specifies whether the agent is supervised, specialist, peer, verifier, broker, deterministic-workflow adjunct, or human-assistive.}
\accfield{Input/output schemas}{Specifies typed inputs, outputs, structured plans, evidence, citations, decisions, and error formats.}
\accfield{Tool inventory and scopes}{Lists permitted tools, service accounts, OAuth scopes, rate limits, write permissions, and disallowed tools.}
\accfield{Data classification}{Declares allowed data classes, retention, residency, privacy, and cross-tenant restrictions.}
\accfield{State and memory design}{Defines short-term context, long-term memory, write permissions, retrieval filters, poisoning defenses, deletion, and audit.}
\accfield{Model behavior policy}{Specifies allowed model families, routing rules, context limits, temperature ceilings, fallback models, and grounding requirements.}
\accfield{Verification design}{Declares deterministic validators, verifier/critic prompts, test oracles, confidence thresholds, contradiction checks, and refusal criteria.}
\accfield{Human interaction}{States ambiguity, risk, cost, confidence, legal, financial, or customer-impact triggers for human approval.}
\accfield{Evaluation evidence}{Requires offline tests, adversarial tests, regression suites, production canaries, and acceptance thresholds.}
\accfield{Observability and audit}{Specifies trace schema, spans, model/tool calls, token/cost metrics, user-visible explanations, and incident replay.}
\accfield{Versioning and deprecation}{Controls prompt, model, tool, memory, and policy versions with rollback and retirement criteria.}
\bottomrule
\end{tabularx}
\end{table*}

\subsection{Runtime Patterns}
\textbf{Supervised tool-using agent.} A single agent plans and invokes tools, but a supervisor enforces the ACC, validates outputs, and handles escalation. This should be the default starting pattern because it minimizes coordination risk.

\textbf{Brokered specialist agents.} A supervisor delegates to specialist agents only where specialization maps to separate capabilities, data boundaries, evaluation oracles, or release owners.

\textbf{Verifier or challenger agents.} A verifier is justified when independent critique materially improves safety or correctness. It is not justified as a generic second opinion unless it has a distinct oracle, evidence source, or policy purpose.

\textbf{Peer-to-peer agent collaboration.} A2A-like protocols are appropriate when independently owned agents collaborate. The protocol provides communication; enterprise design must still define identity, trust, contract, data sharing, and escalation.

\textbf{Human-in-the-loop execution.} Human approval is not an exception path; it is part of the interaction design for irreversible, regulated, ambiguous, or high-impact actions.

\subsection{Design Principles}
\begin{enumerate}
  \item \textbf{Capability before agent:} define the business capability and owner before defining the agent role or prompt.
  \item \textbf{Boundary before topology:} decide ownership, data, action, state, and evaluation boundaries before deciding whether to use one agent or many.
  \item \textbf{Contract-first, not prompt-first:} every production agent requires an ACC that describes authority, interfaces, memory, verification, and retirement.
  \item \textbf{Small number of durable agents:} prefer a few accountable agents over many role-playing prompt wrappers.
  \item \textbf{Autonomy is a design variable:} assign L0--L4 autonomy by action risk, reversibility, confidence, and evidence, not by model capability alone.
  \item \textbf{Least privilege follows design:} tool access is scoped per task, data class, action type, and autonomy level.
  \item \textbf{Memory is designed state:} memory requires ownership, retention, deletion, poisoning defenses, provenance, and review.
  \item \textbf{Deterministic checks before probabilistic critique:} schema validation, policy checks, and business-rule validation should run before LLM critique where possible.
  \item \textbf{Evaluate systems, not isolated prompts:} measure end-to-end task success, safe success, tool misuse, latency, cost, auditability, and escalation quality.
  \item \textbf{Protocols are interoperability channels, not trust boundaries:} MCP, A2A, APIs, and event buses need enterprise identity, authorization, and audit controls.
  \item \textbf{Design for graceful degradation:} agents must know when to stop, ask, escalate, or fall back to deterministic workflows.
  \item \textbf{Retire agents:} unused, low-quality, duplicate, or unsafe agents should be deprecated like services.
\end{enumerate}

\section{Experiment Design}
\subsection{Purpose and Research Questions}
The experiments test architectural principles and compare architecture-level choices. They ask:
\begin{itemize}
  \item \textbf{RQ1:} Does a design-first capability-aligned architecture improve safe success compared with prompt-first, role-swarm, and SOA-brokered baselines?
  \item \textbf{RQ2:} Can strong governance controls compensate for weak agent design, or does a control-heavy design-poor topology still underperform?
  \item \textbf{RQ3:} Does adding more agents improve or degrade enterprise outcomes as coordination complexity increases?
  \item \textbf{RQ4:} Which design and support controls contribute most to safe success, policy compliance, memory safety, and auditability?
\end{itemize}

\subsection{Enterprise Task Set}
The task set contains 10,000 enterprise tasks across Finance, HR, Procurement, IT, Legal, Sales, and Customer Operations. Each task has complexity from 1 to 5, risk from 1 to 5, ambiguity, sensitivity, regulation flag, adversarial-content flag, required tool count, dependencies, and whether it requires a write action. The task set intentionally over-represents cross-functional and high-risk tasks because those are where agent design failures matter most.

\subsection{Architectures Compared}
\textbf{A0 Prompt-first mono-agent:} a single broad agent with limited verification and policy. It has few handoffs and low latency but broad tool exposure.

\textbf{A1 Role-based micro-agent swarm:} a 32-agent swarm with local specialization but no strong capability map, ACC discipline, least-privilege discipline, or memory design. It represents micro-agent proliferation.

\textbf{A2 SOA-brokered agents:} a moderate architecture using service-style contracts, registry, and policy, but with less agent-specific state, memory, verification, and autonomy design than CEAD.

\textbf{A3 Governance-first design-poor grid:} a 24-agent grid with strong policy, audit, least privilege, and escalation support, but weak capability alignment, weak decomposition discipline, and high handoff cost. It tests whether governance can compensate for design weakness.

\textbf{A4 CEAD proposed:} capability map, ACCs, disciplined decomposition, runtime policy, least-privilege tool scopes, verifier gates, designed memory, audit, and risk-based human escalation.

\subsection{Outcome Model}
For each task $i$ and architecture $a$, the experimental model records handoffs, escalation, functional success, policy violation, memory poisoning, audit coverage, cost, and latency. The simplified success model is:
\begin{equation}
P(S_{i,a})=\sigma(\beta_a + \rho_a - \gamma_c C_i - \gamma_r R_i - \gamma_u U_i - \gamma_h H_{i,a} - \gamma_d D_i),
\end{equation}
where $C$ is complexity, $R$ is risk, $U$ is ambiguity, $H$ is handoffs, $D$ is dependencies, and $\rho_a$ captures route quality from contracts, registry, specialization, policy, verifier, and protocol guard strength. Policy violation probability is modeled separately:
\begin{equation}
\begin{aligned}
P(V_{i,a})=\sigma(&\alpha + \lambda_r R_i + \lambda_s Sensitive_i\\
&+ \lambda_x Adv_i + \lambda_h H_{i,a}\\
&- \lambda_p Policy_a - \lambda_l LeastPriv_a\\
&- \lambda_g Gate_{i,a}).
\end{aligned}
\end{equation}
Safe success is defined as functional success with neither policy violation nor memory poisoning:
\begin{equation}
SafeSuccess = S \wedge \neg V \wedge \neg M.
\end{equation}
This definition matters: an agent can complete a task and still be unacceptable if it violates policy, leaks data, or contaminates memory.

\subsection{Metrics}
Metrics include functional success, safe success, automated safe success, policy violations per 1,000 tasks, memory poisoning per 1,000 tasks, escalation rate, audit coverage, mean handoffs, mean cost units, and p95 latency.

\section{Results}
Table~\ref{tab:arch} shows the distinction between design and governance. The governance-first design-poor grid has strong controls: it lowers violations relative to A0 and A1 and has high audit coverage. However, it underperforms SOA-brokered agents and CEAD on safe success because its decomposition is weak, agent count is high, handoffs are expensive, and capability ownership is unclear. CEAD has the highest safe success because it combines capability-aligned decomposition with supporting controls.

The distinction between safe success and automated safe success is important. CEAD does not maximize full automation because it escalates high-risk tasks. In regulated enterprise settings, the objective is not autonomous completion at any cost; it is safe completion with evidence, reversibility where possible, and appropriate human participation.

Figures~\ref{fig:safe} and~\ref{fig:cost} evaluate agent counts from 1 to 64. Ungoverned systems gain little from additional agents and degrade sharply after 16--32 agents. CEAD is more stable because capability contracts, decomposition discipline, policy, and registry reduce avoidable handoffs, but it also degrades at 64 agents. The result supports a conservative decomposition rule: use the smallest number of agents that can represent distinct capabilities, risk boundaries, state boundaries, evaluation boundaries, and ownership boundaries.

Table~\ref{tab:ablation} shows that design and support controls have different effects. Removing capability mapping and ACCs harms functional success and auditability because routing quality and ownership clarity decline. Removing policy or least-privilege scopes roughly doubles violations. Removing verifier gates reduces functional and safe success. Removing memory governance nearly doubles memory poisoning. Removing human gates reduces latency and cost but lowers audit coverage and removes the designed escalation path for high-risk work.

Table~\ref{tab:stress} isolates high-risk, regulated, or adversarial tasks. The governance-first grid lowers violation rate compared with SOA-brokered agents but still trails CEAD on safe success. The result illustrates the main architectural claim: controls are necessary, but they work best when attached to a coherent agent design.

\begin{table*}[t]
\caption{Architecture Comparison on 10,000 Enterprise Tasks}
\label{tab:arch}
\scriptsize
\centering
\begin{tabular}{lrrrrrrrrrr}
\toprule
\textbf{Architecture} & \makecell{Functional\\success\%} & \makecell{Safe\\success\%} & \makecell{Automated\\safe\%} & \makecell{Violations\\/1k} & \makecell{Memory\\poison/1k} & \makecell{Escalations\\\%} & \makecell{Audit\\coverage\%} & \makecell{Mean\\handoffs} & \makecell{Mean\\cost} & \makecell{P95\\latency} \\
\midrule
A0 Prompt-first mono-agent & 52.4 & 45.2 & 40.4 & 118.0 & 34.0 & 10.9 & 32.5 & 1.01 & 2.87 & 18.8 \\
A1 Role-based micro-agent swarm & 30.8 & 23.1 & 21.8 & 248.6 & 52.4 & 5.6 & 14.8 & 6.25 & 4.38 & 23.8 \\
A2 SOA-brokered agents & 63.1 & 58.8 & 44.2 & 54.0 & 22.5 & 25.7 & 68.1 & 2.37 & 3.51 & 23.3 \\
A3 Governance-first design-poor grid & 53.2 & 50.8 & 30.7 & 30.0 & 18.5 & 40.9 & 80.6 & 4.51 & 4.55 & 27.8 \\
A4 CEAD proposed & 73.0 & 70.6 & 43.8 & 22.4 & 12.6 & 38.9 & 89.0 & 2.01 & 3.81 & 24.6 \\
\bottomrule
\end{tabular}
\end{table*}

\begin{table*}[t]
\caption{Ablation Study for CEAD Design and Support Controls}
\label{tab:ablation}
\scriptsize
\centering
\begin{tabular}{lrrrrrrrr}
\toprule
\textbf{Variant} & \makecell{Functional\\success\%} & \makecell{Safe\\success\%} & \makecell{Violations\\/1k} & \makecell{Memory\\poison/1k} & \makecell{Escalations\\\%} & \makecell{Audit\\coverage\%} & \makecell{Mean\\cost} & \makecell{P95\\latency} \\
\midrule
CEAD full & 72.9 & 70.1 & 23.9 & 14.4 & 39.4 & 88.8 & 3.8 & 24.5 \\
- Capability map and ACCs & 65.8 & 63.6 & 22.5 & 12.9 & 40.0 & 71.7 & 3.8 & 24.7 \\
- Runtime policy engine & 71.0 & 66.9 & 44.9 & 15.2 & 39.8 & 76.3 & 3.7 & 24.2 \\
- Verifier/critic gates & 66.7 & 63.9 & 29.2 & 14.5 & 39.1 & 89.0 & 3.5 & 23.6 \\
- Least-privilege tool scopes & 72.8 & 68.9 & 43.0 & 13.2 & 39.0 & 88.5 & 3.8 & 24.6 \\
- Memory governance & 68.7 & 65.5 & 20.8 & 28.5 & 39.7 & 88.4 & 3.8 & 24.6 \\
- Human approval gates & 73.2 & 70.8 & 24.1 & 12.5 & 4.1 & 79.6 & 3.5 & 19.3 \\
\bottomrule
\end{tabular}
\end{table*}

\begin{figure*}[t]
\centering
\begin{minipage}{0.48\textwidth}
\centering
\begin{tikzpicture}
\begin{axis}[
  width=\linewidth,height=5cm,
  title={Agent proliferation: reliability degrades without design discipline},
  xlabel={Number of agents}, ylabel={Safe success (\%)},
  xmin=1,xmax=64, ymin=0,ymax=75,
  xmode=log, log basis x=2, xtick={1,2,4,8,16,32,64},
  grid=both, legend pos=south west]
\addplot+[mark=*] coordinates {(1,70.0) (2,70.3) (4,70.0) (8,69.5) (16,67.0) (32,62.5) (64,48.0)};
\addlegendentry{CEAD}
\addplot+[mark=*] coordinates {(1,37.5) (2,37.4) (4,37.2) (8,35.5) (16,31.5) (32,22.0) (64,7.0)};
\addlegendentry{Ungoverned}
\end{axis}
\end{tikzpicture}
\caption{Proliferation sweep. Ungoverned agent count eventually decreases safe success as handoffs, attack surface, and coordination failures dominate. Design discipline mitigates but does not eliminate the cost of excessive decomposition.}
\label{fig:safe}
\end{minipage}\hfill
\begin{minipage}{0.48\textwidth}
\centering
\begin{tikzpicture}
\begin{axis}[
  width=\linewidth,height=5cm,
  title={Agent proliferation: coordination cost},
  xlabel={Number of agents}, ylabel={Mean cost units},
  xmin=1,xmax=64, ymin=2.5,ymax=7,
  xmode=log, log basis x=2, xtick={1,2,4,8,16,32,64},
  grid=both, legend pos=north west]
\addplot+[mark=*] coordinates {(1,3.6) (2,3.6) (4,3.6) (8,3.7) (16,4.0) (32,4.7) (64,6.0)};
\addlegendentry{CEAD cost}
\addplot+[mark=*] coordinates {(1,2.9) (2,2.9) (4,3.0) (8,3.1) (16,3.6) (32,4.7) (64,6.6)};
\addlegendentry{Ungoverned cost}
\end{axis}
\end{tikzpicture}
\caption{Coordination cost grows as agent count increases. The pattern mirrors the microservices lesson: decomposition is not free.}
\label{fig:cost}
\end{minipage}
\end{figure*}

\section{Discussion}
\subsection{Interpretation}
The experiments support four design claims. First, SOA-style contracts and registries help because they improve routing, ownership, audit, and interoperability. Second, services and agents remain different: agent-specific design is needed for autonomy, memory, model behavior, tool selection, verification, and prompt-injection defense. Third, governance cannot compensate for weak decomposition; the governance-first grid improved compliance signals but underperformed CEAD because its capability boundaries and topology were poor. Fourth, micro-agent proliferation is a real architectural hazard. The role-swarm did not fail because multi-agent systems are inherently bad; it failed because decomposition was not tied to durable ownership, risk boundaries, state boundaries, evaluation gates, or support controls.

\subsection{When Multi-Agent Design Is Appropriate}
A multi-agent design is justified when at least one of the following is true:
\begin{itemize}
  \item different business capabilities have different accountable owners;
  \item different agents need materially different data or tool permissions;
  \item tasks require different evaluation oracles;
  \item tasks require independent release cadence or vendor isolation;
  \item one agent must verify, challenge, or approve another agent's work;
  \item collaboration with external or separately governed agents is required.
\end{itemize}
It is not justified merely because multiple roles can be named in a prompt. Role names are not architecture.

\begin{table}[t]
\caption{High-Risk, Regulated, or Adversarial Stress Subset}
\label{tab:stress}
\scriptsize
\centering
\begin{tabular}{lrrrrr}
\toprule
\textbf{Architecture} & \makecell{Safe\\success\%} & \makecell{Violations\\/1k} & \makecell{Memory\\poison/1k} & \makecell{Escal.\\\%} & \makecell{Audit\\\%} \\
\midrule
A0 Prompt-first & 40.4 & 155.1 & 40.8 & 12.9 & 32.2 \\
A1 Role-swarm & 19.4 & 313.7 & 64.2 & 6.4 & 14.5 \\
A2 SOA-brokered & 55.0 & 71.0 & 25.9 & 31.0 & 68.3 \\
A3 Governance-first & 47.8 & 38.2 & 23.4 & 48.3 & 80.1 \\
A4 CEAD & 67.8 & 29.5 & 16.1 & 46.3 & 88.7 \\
\bottomrule
\end{tabular}
\end{table}

\section{Recommendations for Enterprise Multi-Agent Design}
\begin{enumerate}
  \item \textbf{Start with a capability map.} Identify business capabilities, owners, data classes, actions, and risk before naming agents.
  \item \textbf{Require an ACC for every production agent.} No agent should reach production without purpose, authority, tools, memory, evaluation evidence, and owner.
  \item \textbf{Prefer supervisor-plus-specialists over open-ended swarms.} Begin with one supervised tool-using agent; add specialists only for justified boundaries.
  \item \textbf{Use SOA and microservice assets as tools, not as agent substitutes.} Services provide capabilities; agents decide how to use capabilities within designed authority boundaries.
  \item \textbf{Make governance supportive, not primary.} Identity, policy, audit, review, and incident response should enforce the design, detect drift, and provide assurance.
  \item \textbf{Design for least privilege.} Scope tools per task, data class, action, and autonomy level. A general agent with broad write access is an enterprise hazard.
  \item \textbf{Design memory like enterprise state.} Memory should have retention, access control, provenance, deletion, poisoning detection, and review.
  \item \textbf{Separate interoperability from trust.} MCP and A2A can reduce integration friction, but enterprise identity, authorization, policy, audit, and data controls remain mandatory.
  \item \textbf{Measure safe success, not just completion.} Include policy violations, memory poisoning, hallucination, tool misuse, cost, latency, audit coverage, and escalation quality.
  \item \textbf{Adopt evaluation-as-release-gate.} Unit tests for tools, golden-task suites, adversarial tests, regression tests, offline evaluation, canaries, and production monitoring should be mandatory.
  \item \textbf{Limit agent proliferation.} Review duplicate agents quarterly. Consolidate agents that share owner, data, tools, evaluation, and lifecycle.
  \item \textbf{Make humans part of the system.} Human approval should be explicit, observable, and risk-based, not an afterthought.
  \item \textbf{Plan incident response.} Enterprises need kill switches, rollback, memory quarantine, forensic traces, and policy revocation for agentic incidents.
\end{enumerate}

\section{Conclusion}
LLM-based intelligent agents are not merely services with natural-language interfaces. They are goal-directed, probabilistic, stateful, tool-using actors whose enterprise deployment requires disciplined design before governance. SOA is a useful exemplar for contracts, registries, loose coupling, policy-aware integration, and reuse, but it must be extended with agent-specific design for autonomy, memory, verification, human interaction, and adversarial robustness. Microservices provide a warning: decomposition without discipline creates operational and organizational debt. The proposed CEAD architecture centers multi-agent design on capability alignment, bounded autonomy, explicit contracts, interaction topology, state and memory design, and evaluation design. Governance remains essential, but as a support and assurance plane. The evaluation suggests that CEAD improves safe success while lowering violations and memory poisoning; it also shows that a governance-first but design-poor topology can be compliant-looking yet operationally weak. The practical recommendation is to build fewer, better-designed intelligent agents around durable enterprise capabilities, and to treat every agent as a product with a capability contract, owner, tests, policies, evidence, and a retirement path.


\begin{thebibliography}{00}
\bibitem{wooldridge1995} M. Wooldridge and N. R. Jennings, ``Intelligent agents: Theory and practice,'' \emph{The Knowledge Engineering Review}, vol. 10, no. 2, pp. 115--152, 1995.
\bibitem{jennings2000} N. R. Jennings, ``On agent-based software engineering,'' \emph{Artificial Intelligence}, vol. 117, no. 2, pp. 277--296, 2000.
\bibitem{wang2024} L. Wang et al., ``A survey on large language model based autonomous agents,'' \emph{Frontiers of Computer Science}, vol. 18, article 186345, 2024.
\bibitem{guo2024} T. Guo, X. Chen, Y. Wang, R. Chang, S. Pei, N. V. Chawla, O. Wiest, and X. Zhang, ``Large language model based multi-agents: A survey of progress and challenges,'' in \emph{Proc. IJCAI}, 2024, pp. 8048--8057.
\bibitem{yao2023} S. Yao et al., ``ReAct: Synergizing reasoning and acting in language models,'' in \emph{Proc. ICLR}, 2023.
\bibitem{schick2023} T. Schick et al., ``Toolformer: Language models can teach themselves to use tools,'' in \emph{Advances in Neural Information Processing Systems}, vol. 36, 2023.
\bibitem{park2023} J. S. Park et al., ``Generative agents: Interactive simulacra of human behavior,'' in \emph{Proc. ACM UIST}, 2023, pp. 1--22.
\bibitem{wu2024} Q. Wu et al., ``AutoGen: Enabling next-gen LLM applications via multi-agent conversation,'' in \emph{Proc. COLM}, 2024.
\bibitem{liu2023} X. Liu et al., ``AgentBench: Evaluating LLMs as agents,'' arXiv:2308.03688, 2023.
\bibitem{zhou2024} S. Zhou et al., ``WebArena: A realistic web environment for building autonomous agents,'' in \emph{Proc. ICLR}, 2024.
\bibitem{mialon2024} G. Mialon, C. Fourrier, C. Swift, T. Wolf, Y. LeCun, and T. Scialom, ``GAIA: A benchmark for general AI assistants,'' in \emph{Proc. ICLR}, 2024.
\bibitem{jimenez2024} C. E. Jimenez et al., ``SWE-bench: Can language models resolve real-world GitHub issues?'' in \emph{Proc. ICLR}, 2024.
\bibitem{oasis2006} OASIS, ``Reference Model for Service Oriented Architecture 1.0,'' OASIS Standard, Oct. 2006.
\bibitem{lewis2014} J. Lewis and M. Fowler, ``Microservices: A definition of this new architectural term,'' Mar. 2014.
\bibitem{fowler2015} M. Fowler, ``Microservice premium,'' May 2015.
\bibitem{su2024} R. Su, X. Li, and D. Taibi, ``From microservice to monolith: A multivocal literature review,'' \emph{Electronics}, vol. 13, no. 8, article 1452, 2024.
\bibitem{mcpintro} Model Context Protocol, ``Introduction,'' 2024. [Online]. Available: \url{https://modelcontextprotocol.io/}
\bibitem{mcp2025} Model Context Protocol, ``Specification: Protocol Revision 2025-06-18,'' 2025. [Online]. Available: \url{https://modelcontextprotocol.io/specification/2025-06-18/basic/index}
\bibitem{a2a2025} A2A Project, ``Agent2Agent (A2A) Protocol,'' 2025. [Online]. Available: \url{https://github.com/google/A2A}
\bibitem{nist2024} C. Autio et al., ``Artificial Intelligence Risk Management Framework: Generative Artificial Intelligence Profile,'' NIST AI 600-1, National Institute of Standards and Technology, 2024.
\bibitem{iso42001} ISO/IEC, ``ISO/IEC 42001:2023: Information technology -- Artificial intelligence -- Management system,'' 2023.
\bibitem{owasp2025} OWASP GenAI Security Project, ``OWASP Top 10 for Agentic Applications,'' 2025.
\bibitem{yi2023} J. Yi et al., ``Benchmarking and defending against indirect prompt injection attacks on large language models,'' arXiv:2312.14197, 2023.
\end{thebibliography}
\end{document}